\documentclass[12pt,thmsa]{article}
\usepackage{sw20lart}

\input tcilatex
\QQQ{Language}{
British English
}

\begin{document}

\section{Introduction}

In a series of papers, Mweene has developed a generalized description of
angular momentum which contains the standard results in a certain limit. He
has thereby obtained new generalized expressions for the operators and
eigenvectors for spin 1/2[1-4], spin 1[5], spin 3/2[6], spin 2[7] and spin
5/2[8]. Applying this approach to angular momentum addition, he has shown
how the standard results for various states that arise from combining two
values of angular momentum come about from a consideration of probability
amplitudes for measurements on these systems. This had led to generalized
results for angular momentum addition which also reduce to the standard
results in an appropriate limit[9-10].

In a further development of the approach, Mweene has shown that the usual
spherical harmonics are just special forms of more generalized quantities
and he has obtained the generalized spherical harmonics for the case $l=1$%
[11]. In this paper, we give the generalized spherical harmonics for $l=2$.

This paper is organized as follows. This Introduction is followed in Section
2 by a brief review of the theory underlying the work. Section 3 contains
the derivation of the generalized spherical harmonics and their probability
amplitudes. Section 4 is a discussion of some of the properties of these
quantities. The Discussion and Conclusion in Section 5 closes the paper.

\section{Theoretical Background}

This work is inspired by the interpretation of quantum mechanics due to
Land\'e[12-15]. According to Land\'e, if a quantum system possesses three
sets of observables $A$, $B$ and $C$ with respective eigenvalue spectra $A_1$%
, $A_2,$...,$A_N$, $B_1$, $B_2$, ...,$B_N$ and $C_1$, $C_2$, ....,$C_N$ -
where $N$ is the multiplicity of each spectrum, which is necessarily the
same for each observable[13] - then three sets of probability amplitudes can
be defined for the system. One set, which is denoted by $\psi (A_i,C_j),$
relates to measurement of the observable $C$ if the system is in a state
corresponding to eigenvalues of the observable $A$: thus $\left| \psi
(A_i;C_j)\right| ^2$ gives the probability for obtaining the value $C_j$ if
the initial state corresponds to the eigenvalue $A_i$ of the observable $A$.
The second set $\chi (A_i;B_j)$ relates to measurement of $B$ when the
system is in a state corresponding to the eigenvalue $A$. Finally, the set $%
\phi (B_i;C_j)$ describes measurements of $C$ when the initial state belongs
to the observable $B$. Since the three sets of probability amplitudes belong
to one system, they are interdependent, and the law of interdependence
is[12,13] 
\begin{equation}
\psi (A_i;C_j)=\sum_j\chi (A_i;B_j)\phi (B_j;C_n)  \label{on1}
\end{equation}

Another aspect of Land\'e's interpretation of quantum mechanics is that
every eigenfunction or wave function of a quantum system is first and
foremost a probability amplitude and that every such probability amplitude
connects two well-defined states - one corresponding to the state in which
the system is before a measurement, and the other to the state that comes
about as a result of the measurement[12,13]. It is therefore always possible
to identify an initial and a final state for any wave function or
eigenfunction.

Mweene has argued that for an eigenfunction resulting from solution of a
differential eigenvalue equation, the initial state corresponds to the
eigenvalue while the final state corresponds to the eigenvalue defined by
the continuous variable in terms of which the differential operator is
defined[16]. For the Schr\"odinger equation, the eigenfunctions $\psi
_{E_i}(x)$ should really be written as $\psi (E_i;x)$ to emphasize that the
initial state in the probability amplitude corresponds to the eigenvalue $%
E_i $ while the final state corresponds to the eigenvalue $x$. Another
example comes from the solution of Legendre's equation. The spherical
harmonics $Y_{lm}(\theta ,\varphi )$ should really be written as $%
Y(l,m;\theta ,\varphi )$ to emphasize that in this case the initial state is
defined by the eigenvalues $m\hbar $ and $l(l+1)\hbar ^2$ while the final
state corresponds to the angular position $(\theta ,\varphi ).$ Since $%
m\hbar $ is an angular momentum projection, it must be defined with respect
to some axis. Owing to the absence of another set of angles in the
expressions of the spherical harmonics which could define this direction, it
must be the $z$ axis[11]. But since an axis of quantization can be chosen
arbitrarily, it is possible to define spherical harmonics with respect to
any other direction as the axis of initial quantization. The functions
resulting from this are the generalized spherical harmonics and have already
been worked out for the case $l=1$. In this work, we obtain them for $l=2$.

\section{Generalized Spherical Harmonics}

\subsection{Probability Amplitudes}

The generalized spherical harmonics connect states of angular momentum
projection in the arbitrary direction $\widehat{\mathbf{a}}$ defined by the
polar angles $(\theta ^{\prime },\varphi ^{\prime })$ to states of the
angular position $(\theta ,\varphi ).$ We denote them by $Y(l,m^{(\widehat{%
\mathbf{a}})};\theta ,\varphi )$ . To derive them, we use the probability
addition law Eq. (\ref{on1}). We start off by writing 
\begin{equation}
Y(l,m^{(\widehat{\mathbf{a}})};\theta ,\varphi )=\sum_j\chi (l,m^{(\widehat{%
\mathbf{a}})};B_j)\phi (B_j;\theta ,\varphi )  \label{th34}
\end{equation}
If we choose the observable $B$ carefully, we should find that both the
probability amplitudes $\chi (l,m^{(\widehat{\mathbf{a}})};B_j)$ and $\phi
(B_j;\theta ,\varphi )$ are known. If $B$ is chosen to be the spin
projection with respect to the $z$ direction, it is found that 
\begin{equation}
\phi (B_j;\theta ,\varphi )=Y(l,m^{(\widehat{\mathbf{k}})};\theta ,\varphi )
\label{th34a}
\end{equation}
are the standard spherical harmonics, while $\chi (l,m_i^{(\widehat{\mathbf{a%
}})};l,m_f^{(\widehat{\mathbf{k}})})$ are just spin probability amplitudes
connecting states such that the initial one corresponds to the spin
projection being $m_i\hbar $ in the direction $\widehat{\mathbf{a}}$ while
the final state corresponds to the spin projection being $m_f\hbar $ along
the $z $ axis. These have already been worked out[7] and are as given below.
If the initial spin projection in the direction $\widehat{\mathbf{a}}$ is $%
2\hbar $, these probability amplitudes are

\begin{equation}
\chi (2,2^{(\widehat{\mathbf{a}})};2,2^{(\widehat{\mathbf{k}})})=\cos ^4%
\frac{\theta ^{\prime }}2e^{-2i\varphi ^{\prime }}  \label{fi53}
\end{equation}
\begin{equation}
\chi (2,2^{(\widehat{\mathbf{a}})};2,1^{(\widehat{\mathbf{k}})})=2\sin \frac{%
\theta ^{\prime }}2\cos ^3\frac{\theta ^{\prime }}2e^{-i\varphi ^{\prime }}
\label{fi54}
\end{equation}
\begin{equation}
\chi (2,2^{(\widehat{\mathbf{a}})};2,0^{(\widehat{\mathbf{k}})})=\sqrt{6}%
\sin ^2\frac{\theta ^{\prime }}2\cos ^2\frac{\theta ^{\prime }}2
\label{fi55}
\end{equation}
\begin{equation}
\chi (2,2^{(\widehat{\mathbf{a}})};2,(-1)^{(\widehat{\mathbf{k}})})=2\sin ^3%
\frac{\theta ^{\prime }}2\cos \frac{\theta ^{\prime }}2e^{i\varphi ^{\prime
}}  \label{fi56}
\end{equation}
and 
\begin{equation}
\chi (2,2^{(\widehat{\mathbf{a}})};2,(-2)^{(\widehat{\mathbf{k}})})=\sin ^4%
\frac{\theta ^{\prime }}2e^{2i\varphi ^{\prime }}  \label{fi57}
\end{equation}

If the initial spin projection in the direction $\widehat{\mathbf{a}}$ is $%
\hbar ,$ the probability amplitudes are 
\begin{equation}
\chi (2,1^{(\widehat{\mathbf{a}})};2,2^{(\widehat{\mathbf{k}})})=2\sin \frac{%
\theta ^{\prime }}2\cos ^3\frac{\theta ^{\prime }}2e^{-2i\varphi ^{\prime }}
\label{fi58}
\end{equation}

\begin{equation}
\chi (2,1^{(\widehat{\mathbf{a}})};2,1^{(\widehat{\mathbf{k}})})=-(3\sin ^2%
\frac{\theta ^{\prime }}2-\cos ^2\frac{\theta ^{\prime }}2)\cos ^2\frac{%
\theta ^{\prime }}2e^{-i\varphi ^{\prime }}  \label{fi59}
\end{equation}
\begin{equation}
\chi (2,1^{(\widehat{\mathbf{a}})};2,0^{(\widehat{\mathbf{k}})})=-\sqrt{6}%
\cos \frac{\theta ^{\prime }}2\sin \frac{\theta ^{\prime }}2\cos \theta
^{\prime }  \label{si60}
\end{equation}
\begin{equation}
\chi (2,1^{(\widehat{\mathbf{a}})};2,(-1)^{(\widehat{\mathbf{k}})})=(3\cos ^2%
\frac{\theta ^{\prime }}2-\sin ^2\frac{\theta ^{\prime }}2)\sin ^2\frac{%
\theta ^{\prime }}2e^{i\varphi ^{\prime }}  \label{si61}
\end{equation}

and 
\begin{equation}
\chi (2,1^{(\widehat{\mathbf{a}})};2,(-2)^{(\widehat{\mathbf{k}})})=-2\sin ^3%
\frac{\theta ^{\prime }}2\cos \frac{\theta ^{\prime }}2e^{2i\varphi ^{\prime
}}  \label{si62}
\end{equation}

If the initial spin projection in the direction $\widehat{\mathbf{a}}$ is $%
0, $the probability amplitudes are 
\begin{equation}
\chi (2,0^{(\widehat{\mathbf{a}})};2,2^{(\widehat{\mathbf{k}})})=\sqrt{6}%
\sin ^2\frac{\theta ^{\prime }}2\cos ^2\frac{\theta ^{\prime }}%
2e^{-2i\varphi ^{\prime }}  \label{si63}
\end{equation}
\begin{equation}
\chi (2,0^{(\widehat{\mathbf{a}})};2,1^{(\widehat{\mathbf{k}})})=-\sqrt{6}%
\sin \frac{\theta ^{\prime }}2\cos \frac{\theta ^{\prime }}2\cos \theta
^{\prime }e^{-i\varphi ^{\prime }}  \label{si64}
\end{equation}
\begin{equation}
\chi (2,0^{(\widehat{\mathbf{a}})};2,0^{(\widehat{\mathbf{k}})})=\frac
12(2\cos ^2\theta ^{\prime }-\sin ^2\theta ^{\prime })  \label{si65}
\end{equation}
\begin{equation}
\chi (2,0^{(\widehat{\mathbf{a}})};2,(-1)^{(\widehat{\mathbf{k}})})=\sqrt{6}%
\sin \frac{\theta ^{\prime }}2\cos \frac{\theta ^{\prime }}2\cos \theta
^{\prime }e^{i\varphi ^{\prime }}  \label{si66}
\end{equation}
and 
\begin{equation}
\chi (2,0^{(\widehat{\mathbf{a}})};2,(-2)^{(\widehat{\mathbf{k}})})=\sqrt{6}%
\sin ^2\frac{\theta ^{\prime }}2\cos ^2\frac{\theta ^{\prime }}2e^{2i\varphi
^{\prime }}  \label{si67}
\end{equation}

If the initial spin projection in the direction $\widehat{\mathbf{a}}$ is $%
-\hbar ,$ the probability amplitudes are 
\begin{equation}
\chi (2,(-1)^{(\widehat{\mathbf{a}})};2,2^{(\widehat{\mathbf{k}})})=2\cos 
\frac{\theta ^{\prime }}2\sin ^3\frac{\theta ^{\prime }}2e^{-2i\varphi
^{\prime }}  \label{si68}
\end{equation}
\begin{equation}
\chi (2,(-1)^{(\widehat{\mathbf{a}})};2,1^{(\widehat{\mathbf{k}})})=-(3\cos
^2\frac{\theta ^{\prime }}2-\sin ^2\frac{\theta ^{\prime }}2)\sin ^2\frac{%
\theta ^{\prime }}2e^{-i\varphi ^{\prime }}  \label{si69}
\end{equation}
\begin{equation}
\chi (2,(-1)^{(\widehat{\mathbf{a}})};2,0^{(\widehat{\mathbf{k}})})=\sqrt{6}%
\sin \frac{\theta ^{\prime }}2\cos \frac{\theta ^{\prime }}2\cos \theta
^{\prime }  \label{se70}
\end{equation}
\begin{equation}
\chi (2,(-1)^{(\widehat{\mathbf{a}})};2,(-1)^{(\widehat{\mathbf{k}}%
)})=(3\sin ^2\frac{\theta ^{\prime }}2-\cos ^2\frac{\theta ^{\prime }}2)\cos
^2\frac{\theta ^{\prime }}2e^{i\varphi ^{\prime }}  \label{se71}
\end{equation}
and 
\begin{equation}
\chi (2,(-1)^{(\widehat{\mathbf{a}})};2,(-2)^{(\widehat{\mathbf{k}}%
)})=-2\sin \frac{\theta ^{\prime }}2\cos ^3\frac{\theta ^{\prime }}%
2e^{2i\varphi ^{\prime }}  \label{se72}
\end{equation}

Finally, if the initial spin projection in the direction $\widehat{\mathbf{a}%
}$ is $-2\hbar ,$ the probability amplitudes are

\begin{equation}
\chi (2,(-2)^{(\widehat{\mathbf{a}})};2,2^{(\widehat{\mathbf{k}})})=\sin ^4%
\frac{\theta ^{\prime }}2e^{-2i\varphi ^{\prime }}  \label{se73}
\end{equation}
\begin{equation}
\chi (2,(-2)^{(\widehat{\mathbf{a}})};2,1^{(\widehat{\mathbf{k}})})=-2\cos 
\frac{\theta ^{\prime }}2\sin ^3\frac{\theta ^{\prime }}2e^{-i\varphi
^{\prime }}  \label{se74}
\end{equation}
\begin{equation}
\chi (2,(-2)^{(\widehat{\mathbf{a}})};2,0^{(\widehat{\mathbf{k}})})=\sqrt{6}%
\sin ^2\frac{\theta ^{\prime }}2\cos ^2\frac{\theta ^{\prime }}2
\label{se75}
\end{equation}
\begin{equation}
\chi (2,(-2)^{(\widehat{\mathbf{a}})};2,(-1)^{(\widehat{\mathbf{k}}%
)})=-2\sin \frac{\theta ^{\prime }}2\cos ^3\frac{\theta ^{\prime }}%
2e^{i\varphi ^{\prime }}  \label{se76}
\end{equation}
and 
\begin{equation}
\chi (2,(-2)^{(\widehat{\mathbf{a}})};2,(-2)^{(\widehat{\mathbf{k}})})=\cos
^4\frac{\theta ^{\prime }}2e^{2i\varphi ^{\prime }}  \label{se77}
\end{equation}

The ordinary spherical harmonics $Y_{2m}(\theta ,\varphi )=Y(2,m^{(\widehat{%
\mathbf{k}})};\theta ,\varphi )$ for $l=2$ are

\begin{equation}
Y(2,2^{(\widehat{\mathbf{k}})};\theta ,\varphi )=\sqrt{\frac{15}{32\pi }}%
\sin ^2\theta e^{2i\varphi }  \label{se78}
\end{equation}

\begin{equation}
Y(2,1^{(\widehat{\mathbf{k}})};\theta ,\varphi )=-\sqrt{\frac{15}{8\pi }}%
\sin \theta \cos \theta e^{i\varphi }  \label{se79}
\end{equation}

\begin{equation}
Y(2,0^{(\widehat{\mathbf{k}})};\theta ,\varphi )=\sqrt{\frac 5{16\pi }}%
(3\cos ^2\theta -1)  \label{ei80}
\end{equation}

\begin{equation}
Y(2,(-1)^{(\widehat{\mathbf{k}})};\theta ,\varphi )=\sqrt{\frac{15}{8\pi }}%
\sin \theta \cos \theta e^{-i\varphi }  \label{ei81}
\end{equation}

\begin{equation}
Y(2,(-2)^{(\widehat{\mathbf{k}})};\theta ,\varphi )=\sqrt{\frac{15}{32\pi }}%
\sin ^2\theta e^{-2i\varphi }  \label{ei82}
\end{equation}

Using Eq. (\ref{th34}), the generalized spherical harmonics for $l=2$ are
found to be

\begin{eqnarray}
Y(2,2^{(\mathbf{\hat a})};\theta ,\varphi ) &=&\sqrt{\frac{15}{32\pi }}%
\{\sin ^2\theta (\cos ^4\frac{\theta ^{\prime }}2e^{2i(\varphi -\varphi
^{\prime })}+\sin ^4\frac{\theta ^{\prime }}2e^{-2i(\varphi -\varphi
^{\prime })})  \nonumber  \label{eq52} \\
&&+\sin 2\theta \sin \theta ^{\prime }(-\cos ^2\frac{\theta ^{\prime }}%
2e^{i(\varphi -\varphi ^{\prime })}+\sin ^2\frac{\theta ^{\prime }}%
2e^{-i(\varphi -\varphi ^{\prime })})  \nonumber \\
&&+\frac 12\sin ^2\theta ^{\prime }(3\cos ^2\theta -1)\}  \label{ei83}
\end{eqnarray}

\begin{eqnarray}
Y(2,1^{(\mathbf{\hat a})};\theta ,\varphi ) &=&\sqrt{\frac{15}{32\pi }}%
\{\sin \theta ^{\prime }\sin ^2\theta (\cos ^2\frac{\theta ^{\prime }}%
2e^{2i(\varphi -\varphi ^{\prime })}-\sin ^2\frac{\theta ^{\prime }}%
2e^{-2i(\varphi -\varphi ^{\prime })})  \nonumber \\
&&\ -\sin 2\theta [(3\sin ^2\frac{\theta ^{\prime }}2-\cos ^2\frac{\theta
^{\prime }}2)\cos ^2\frac{\theta ^{\prime }}2e^{i(\varphi -\varphi ^{\prime
})}  \nonumber \\
&&\ +(3\cos ^2\frac{\theta ^{\prime }}2-\sin ^2\frac{\theta ^{\prime }}%
2)\sin ^2\frac{\theta ^{\prime }}2e^{-i(\varphi -\varphi ^{\prime })}] 
\nonumber \\
&&\ -\frac 12\sin 2\theta ^{\prime }(3\cos ^2\theta -1)\}  \label{ei84}
\end{eqnarray}

\begin{eqnarray}
Y(2,0^{(\mathbf{\hat a})};\theta ,\varphi ) &=&\sqrt{\frac{45}{256\pi }}%
\{\sin ^2\theta ^{\prime }\sin ^2\theta (e^{2i(\varphi -\varphi ^{\prime
})}+e^{-2i(\varphi -\varphi ^{\prime })})  \nonumber  \label{eq54} \\
&&+\sin 2\theta \sin 2\theta ^{\prime }(e^{i(\varphi -\varphi ^{\prime
})}+e^{-i(\varphi -\varphi ^{\prime })})  \nonumber \\
&&+\frac 23(3\cos ^2\theta -1)(2\cos ^2\theta ^{\prime }-\sin ^2\theta
^{\prime })\}  \label{ei85}
\end{eqnarray}

\begin{eqnarray}
Y(2,(-1)^{(\mathbf{\hat a})};\theta ,\varphi ) &=&\sqrt{\frac{15}{32\pi }}%
\{\sin \theta ^{\prime }\sin ^2\theta (\sin ^2\frac{\theta ^{\prime }}%
2e^{2i(\varphi -\varphi ^{\prime })}-\cos ^2\frac{\theta ^{\prime }}%
2e^{-2i(\varphi -\varphi ^{\prime })})  \nonumber  \label{eq55} \\
&&\ +\sin 2\theta [(3\cos ^2\frac{\theta ^{\prime }}2-\sin ^2\frac{\theta
^{\prime }}2)\sin ^2\frac{\theta ^{\prime }}2e^{i(\varphi -\varphi ^{\prime
})}  \nonumber \\
&&\ +(3\sin ^2\frac{\theta ^{\prime }}2-\cos ^2\frac{\theta ^{\prime }}%
2)\cos ^2\frac{\theta ^{\prime }}2e^{-i(\varphi -\varphi ^{\prime })}] 
\nonumber \\
&&\ +\frac 12\sin 2\theta ^{\prime }(3\cos ^2\theta -1)\}  \label{ei86}
\end{eqnarray}

\begin{eqnarray}
Y(2,(-2)^{(\mathbf{\hat a})};\theta ,\varphi ) &=&\sqrt{\frac{15}{32\pi }}%
\{\sin ^2\theta (\sin ^4\frac{\theta ^{\prime }}2e^{2i(\varphi -\varphi
^{\prime })}+\cos ^4\frac{\theta ^{\prime }}2e^{-2i(\varphi -\varphi
^{\prime })})  \nonumber  \label{eq56} \\
&&\ +\sin 2\theta \sin \theta ^{\prime }(\sin ^2\frac{\theta ^{\prime }}%
2e^{i(\varphi -\varphi ^{\prime })}-\cos ^2\frac{\theta ^{\prime }}%
2e^{-i(\varphi -\varphi ^{\prime })})  \nonumber \\
&&\ +\frac 12\sin ^2\theta ^{\prime }(3\cos ^2\theta -1)\}  \label{ei87}
\end{eqnarray}

\subsection{Probability Amplitudes for the $x^{\prime }$ Direction}

The results we have presented refer to the direction $\widehat{\mathbf{a}}$
as the direction of initial quantization. We may think of the vector $%
\widehat{\mathbf{a}}$ as defining a new $z$ axis, which we denote by $%
z^{\prime }$, since in the limit $\theta ^{\prime }=\varphi ^{\prime }=0,$
the results corresponding to it reduce to those for the $z$ axis. This $%
z^{\prime }$ axis corresponds to a new coordinate system in which the unit
vector in the $x^{\prime }$ direction is $\widehat{\mathbf{u}}$ and that in
the $y^{\prime }$ direction is $\widehat{\mathbf{v}}$[11]. From the results
for the $\widehat{\mathbf{a}}$ or $z^{\prime }$ axis, we can obtain the
probability amplitudes and probabilities densities for the $x^{\prime }$
direction by applying the transformation $\theta ^{\prime }\rightarrow
\theta ^{\prime }-\pi /2$ to them[3,5]. We are justified in associating the
results so obtained with the $x^{\prime }$ axis since in the limit $\theta
^{\prime }=\varphi ^{\prime }=0, $ they reduce to those for the $x$
direction. When we make these argument changes, $\widehat{\mathbf{a}}$
becomes $\widehat{\mathbf{u}}$. Applying this prescription to the
generalized spherical harmonics, we obtain the results

\begin{eqnarray}
Y(2,2^{(\widehat{\mathbf{u}})};\theta ,\varphi ) &=&\sqrt{\frac{15}{128\pi }}%
\{\frac 12\sin ^2\theta [(1+\sin \theta ^{\prime })^2e^{2i(\varphi -\varphi
^{\prime })}  \nonumber \\
&&+(1-\sin \theta ^{\prime })^2e^{-2i(\varphi -\varphi ^{\prime })}] 
\nonumber \\
&&\ +\sin 2\theta \cos \theta ^{\prime }[(1+\sin \theta ^{\prime
})e^{i(\varphi -\varphi ^{\prime })}  \nonumber \\
&&-(1-\sin \theta ^{\prime })e^{-i(\varphi -\varphi ^{\prime })}]+\cos
^2\theta ^{\prime }(3\cos ^2\theta -1)\}  \label{ni99a}
\end{eqnarray}

\begin{eqnarray}
Y(2,1^{(\widehat{\mathbf{u}})};\theta ,\varphi ) &=&\sqrt{\frac{15}{128\pi }}%
\{-\sin ^2\theta \cos \theta ^{\prime }[(1+\sin \theta ^{\prime
})e^{2i(\varphi -\varphi ^{\prime })}  \nonumber \\
&&-(1-\sin \theta ^{\prime })e^{-2i(\varphi -\varphi ^{\prime })}]  \nonumber
\\
&&\ -\sin 2\theta [(1-2\sin \theta ^{\prime })(1+\sin \theta ^{\prime
})e^{i(\varphi -\varphi ^{\prime })}  \nonumber \\
&&+(1+2\sin \theta ^{\prime })(1-\sin \theta ^{\prime })e^{-i(\varphi
-\varphi ^{\prime })}]  \nonumber \\
&&\ +\sin 2\theta ^{\prime }(3\cos ^2\theta -1)\}  \label{ni99b}
\end{eqnarray}

\begin{eqnarray}
Y(2,0^{(\widehat{\mathbf{u}})};\theta ,\varphi ) &=&\sqrt{\frac{15}{128\pi }}%
\{\cos ^2\theta ^{\prime }\sin ^2\theta (e^{2i(\varphi -\varphi ^{\prime
})}+e^{-2i(\varphi -\varphi ^{\prime })})  \nonumber \\
&&\ -\sin 2\theta \sin 2\theta ^{\prime }[e^{i(\varphi -\varphi ^{\prime
})}+e^{-i(\varphi -\varphi ^{\prime })}]  \nonumber \\
&&\ +\frac 23(2\sin ^2\theta ^{\prime }-\cos ^2\theta ^{\prime })(3\cos
^2\theta -1)\}  \label{ni99c}
\end{eqnarray}

\begin{eqnarray}
Y(2,(-1)^{(\widehat{\mathbf{u}})};\theta ,\varphi ) &=&\sqrt{\frac{15}{%
128\pi }}\{-\sin ^2\theta \cos \theta ^{\prime }[(1-\sin \theta ^{\prime
})e^{2i(\varphi -\varphi ^{\prime })}  \nonumber \\
&&-(1+\sin \theta ^{\prime })e^{-2i(\varphi -\varphi ^{\prime })}]  \nonumber
\\
&&\ +\sin 2\theta [(1+2\sin \theta ^{\prime })(1-\sin \theta ^{\prime
})e^{i(\varphi -\varphi ^{\prime })}  \nonumber \\
&&+(1-2\sin \theta ^{\prime })(1+\sin \theta ^{\prime })e^{-i(\varphi
-\varphi ^{\prime })}]  \nonumber \\
&&\ -\sin 2\theta ^{\prime }(3\cos ^2\theta -1)\}  \label{ni99d}
\end{eqnarray}

\begin{eqnarray}
Y(2,(-2)^{(\widehat{\mathbf{u}})};\theta ,\varphi ) &=&\sqrt{\frac{15}{%
128\pi }}\{\frac 12\sin ^2\theta [(1-\sin \theta ^{\prime })^2e^{2i(\varphi
-\varphi ^{\prime })}  \nonumber \\
&&+(1+\sin \theta ^{\prime })^2e^{-2i(\varphi -\varphi ^{\prime })} 
\nonumber \\
&&\ -\sin 2\theta \cos \theta ^{\prime }[(1-\sin \theta ^{\prime
})e^{i(\varphi -\varphi ^{\prime })}  \nonumber \\
&&-(1+\sin \theta ^{\prime })e^{-i(\varphi -\varphi ^{\prime })}]+\cos
^2\theta ^{\prime }(3\cos ^2\theta -1)\}  \label{ni99e}
\end{eqnarray}

\subsection{Probability Amplitudes for the $y^{\prime }$ Direction}

The prescription for obtaining the probability amplitudes and probability
densities corresponding to $y^{\prime }$ is to set $\theta ^{\prime }=\pi /2$%
, $\varphi ^{\prime }\rightarrow \varphi ^{\prime }-\pi /2$ in the
expressions corresponding to the $z^{\prime }$ direction. As well as
transforming the unit vector $\widehat{\mathbf{a}}$ to the unit vector $%
\widehat{\mathbf{v}}$, this yields the probability amplitudes:

\begin{eqnarray}
Y(2,2^{(\widehat{\mathbf{v}})};\theta ,\varphi ) &=&-\sqrt{\frac{15}{128\pi }%
}\{\frac 12\sin ^2\theta (e^{2i(\varphi -\varphi ^{\prime })}+e^{-2i(\varphi
-\varphi ^{\prime })})  \nonumber \\
&&+i\sin 2\theta [e^{i(\varphi -\varphi ^{\prime })}+e^{-i(\varphi -\varphi
^{\prime })}]-(3\cos ^2\theta -1)\}  \label{ni99k}
\end{eqnarray}

\begin{eqnarray}
Y(2,1^{(\widehat{\mathbf{v}})};\theta ,\varphi ) &=&\sqrt{\frac{15}{128\pi }}%
\{\sin ^2\theta (-e^{2i(\varphi -\varphi ^{\prime })}+e^{-2i(\varphi
-\varphi ^{\prime })})  \nonumber \\
&&\ +i\sin 2\theta [e^{i(\varphi -\varphi ^{\prime })}-e^{-i(\varphi
-\varphi ^{\prime })}]\}  \label{ni99l}
\end{eqnarray}

\begin{eqnarray}
Y(2,0^{(\widehat{\mathbf{v}})};\theta ,\varphi ) &=&-\sqrt{\frac{45}{256\pi }%
}\{\sin ^2\theta (e^{2i(\varphi -\varphi ^{\prime })}+e^{-2i(\varphi
-\varphi ^{\prime })})  \nonumber \\
&&\ \ +\frac 23(3\cos ^2\theta -1)\}  \label{ni99m}
\end{eqnarray}

\begin{eqnarray}
Y(2,(-1)^{(\widehat{\mathbf{v}})};\theta ,\varphi ) &=&\sqrt{\frac{15}{%
128\pi }}\{\sin ^2\theta (-e^{2i(\varphi -\varphi ^{\prime
})}+e^{-2i(\varphi -\varphi ^{\prime })})  \nonumber \\
&&\ \ +i\sin 2\theta [e^{i(\varphi -\varphi ^{\prime })}-e^{-i(\varphi
-\varphi ^{\prime })}]\}  \label{ni99n}
\end{eqnarray}

\begin{eqnarray}
Y(2,(-2)^{(\widehat{\mathbf{v}})};\theta ,\varphi ) &=&\sqrt{\frac{15}{%
128\pi }}\{-\frac 12\sin ^2\theta (e^{2i(\varphi -\varphi ^{\prime
})}+e^{-2i(\varphi -\varphi ^{\prime })})  \nonumber \\
&&\ +i\sin 2\theta [e^{i(\varphi -\varphi ^{\prime })}+e^{-i(\varphi
-\varphi ^{\prime })}]+3\cos ^2\theta -1\}  \nonumber \\
&&  \label{ni99o}
\end{eqnarray}

We emphasize that the unit vectors $\widehat{\mathbf{u}}$, $\widehat{\mathbf{%
v}}$ and $\widehat{\mathbf{a}}$ define a system of mutually orthogonal
coordinate axes.

\section{ General Properties of the Generalized Spherical harmonics}

The generalized quantities presented here reduce to the standard quantities
in the limit $\theta ^{\prime }=\varphi ^{\prime }=0$, which corresponds to
the arbitrary vector $\widehat{\mathbf{a}}$ pointing in the direction of the 
$z$ axis. Thus, in this limit, we get 
\begin{equation}
Y(2,m^{(\mathbf{\hat a})};\theta ,\varphi )\rightarrow Y_{2m}(\theta
,\varphi )  \label{ni99u}
\end{equation}
A property of special interest with regard to the ordinary spherical
harmonics is their behaviour under the parity operation $\mathbf{r}%
\rightarrow -\mathbf{r,}$ a reflection in the origin. Under this operation,
the spherical polar coordinates $(r,\theta ,\varphi )$ transform thus: $%
r\rightarrow r,$ $\theta \rightarrow \pi -\theta ,,\;\phi \rightarrow \phi
+\pi $ . Thus if $\rho $ is the parity operator defined by 
\begin{equation}
\rho \Psi (\mathbf{r})=\Psi (-\mathbf{r}).  \label{hu100}
\end{equation}
then 
\begin{equation}
\rho Y(l,m^{(\widehat{\mathbf{k}})};\theta ,\varphi )=Y(l,m^{(\widehat{%
\mathbf{k}})};\pi -\theta ,\varphi +\pi )  \label{hu101}
\end{equation}
As is well-known however, 
\begin{equation}
Y(l,m^{(\widehat{\mathbf{k}})};\pi -\theta ,\varphi +\pi )=(-1)^lY(l,m^{(%
\widehat{\mathbf{k}})};\theta ,\varphi )  \label{hu102}
\end{equation}
so that $Y(l,m^{(\widehat{\mathbf{k}})};\theta ,\varphi )$ has even parity
if $l $ is even and odd parity if $l$ is odd.

The generalized spherical harmonics have the form 
\begin{equation}
Y(l,m^{(\widehat{\mathbf{a}})};\theta ,\varphi )=\sum_jc_jY(l,m_j^{(\widehat{%
\mathbf{k}})};\theta ,\varphi )  \label{hu103}
\end{equation}
where $c_j=\chi (l,m_i^{(\widehat{\mathbf{a}})};l,m_f^{(\widehat{\mathbf{k}}%
)})$ is a constant with respect to the angles $(\theta ,\varphi ).$ Hence, 
\begin{equation}
\rho Y(l,m^{(\widehat{\mathbf{a}})};\theta ,\varphi )=\sum_jc_j\rho
Y(l,m_j^{(\widehat{\mathbf{k}})};\theta ,\varphi )=(-1)^lY(l,m^{(\widehat{%
\mathbf{a}})};\theta ,\varphi )  \label{hu104}
\end{equation}
Thus, the generalized spherical harmonics have the same parity as the
corresponding standard spherical harmonics.

The generalized spherical harmonics for value of l can be shown to be
orthonormal: 
\begin{equation}
\iint Y^{*}(l,m^{\prime (\widehat{\mathbf{a}})};\theta ,\varphi )Y(l,m^{(%
\widehat{\mathbf{a}})};\theta ,\varphi )d\Omega =\delta _{m^{\prime }m}
\label{hu105}
\end{equation}
Thus, since 
\begin{equation}
Y(l,m^{(\widehat{\mathbf{a}})};\theta ,\varphi )=\sum_j\chi (l,m;l,m_j^{(%
\widehat{\mathbf{k}})})Y(m_j^{(\widehat{\mathbf{k}})};\theta ,\varphi )
\label{hu106}
\end{equation}
and 
\begin{equation}
Y^{*}(l,m^{\prime (\widehat{\mathbf{a}})};\theta ,\varphi )=\sum_{j^{\prime
}}\chi ^{*}(l,m^{\prime (\widehat{\mathbf{a}})};l,m_{j^{\prime }}^{(\widehat{%
\mathbf{k}})})Y^{*}(m_{j^{\prime }}^{(\widehat{\mathbf{k}})};\theta ,\varphi
)  \label{hu107}
\end{equation}
the overlap integral is 
\begin{eqnarray}
I &=&\iint \sum_{j^{\prime }}\chi ^{*}(l,m^{\prime (\widehat{\mathbf{a}}%
)};l,m_{j^{\prime }}^{(\widehat{\mathbf{k}})})Y^{*}(l,m_{j^{\prime }}^{(%
\widehat{\mathbf{k}})};\theta ,\varphi )  \nonumber \\
&&\times \sum_j\chi (l,m^{(\widehat{\mathbf{a}})};l,m_j^{(\widehat{\mathbf{k}%
})})Y(l,m_j^{(\widehat{\mathbf{k}})};\theta ,\varphi )d\Omega \\
&=&\sum_{j^{\prime }}\sum_j\chi ^{*}(l,m^{\prime (\widehat{\mathbf{a}}%
)};l,m_{j^{\prime }}^{(\widehat{\mathbf{k}})})\chi (l,m^{(\widehat{\mathbf{a}%
})};l,m_j^{(\widehat{\mathbf{k}})})  \nonumber \\
&&\times \iint Y^{*}(l,m_{j^{\prime }}^{(\widehat{\mathbf{k}})};\theta
,\varphi )Y(l,m_j^{(\widehat{\mathbf{k}})};\theta ,\varphi )d\Omega \\
&=&\sum_{j^{\prime }}\sum_j\chi ^{*}(l,m^{\prime (\widehat{\mathbf{a}}%
)};l,m_{j^{\prime }}^{(\widehat{\mathbf{k}})})\chi (l,m^{(\widehat{\mathbf{a}%
})};l,m_j^{(\widehat{\mathbf{k}})})\delta _{m_jm_{j^{\prime }}}  \nonumber \\
&=&\sum_j\chi ^{*}(l,m^{\prime (\widehat{\mathbf{a}})};l,m_j^{(\widehat{%
\mathbf{k}})})\chi (l,m;l,m_j^{(\widehat{\mathbf{k}})})  \nonumber \\
&=&\delta _{m^{\prime }m}  \label{hu108}
\end{eqnarray}
In the proof, we have used the result 
\begin{equation}
\sum_j\chi ^{*}(l,m^{\prime (\widehat{\mathbf{a}})};l,m_j^{(\widehat{\mathbf{%
k}})})\chi (l,m;l,m_j^{(\widehat{\mathbf{k}})})=\delta _{m^{\prime }m}
\label{hu109}
\end{equation}
which is just the orthonormality relation for the spin probability
amplitudes.

\section{Discussion and Conclusion}

This work has extended the derivation of the new generalized spherical
harmonics to the case $l=2$. The expressions for the functions have been
derived, as well as the corresponding probability densities for the $%
z^{\prime }$ direction. By means of simple transformations the corresponding
expressions for the $x^{\prime }$ and $y^{\prime }$ directions have been
obtained.

Now, for the case $l=1$, it has been shown that the generalized spherical
harmonics satisfy the eigenvalue equation[11]

\begin{equation}
L_{(\widehat{\mathbf{a}})}Y(1,m^{(\widehat{\mathbf{a}})};\theta ,\varphi
)=m\hbar Y(1,m^{(\widehat{\mathbf{a}})};\theta ,\varphi )  \label{hu110}
\end{equation}
where

\begin{equation}
L_{(\widehat{\mathbf{a}})}=i\hbar \{\sin \theta ^{\prime }\sin (\varphi
-\varphi ^{\prime })\frac \partial {\partial \theta }+[\sin \theta ^{\prime
}\cot \theta \cos (\varphi -\varphi ^{\prime })-\cos \theta ^{\prime }]\frac
\partial {\partial \varphi }\}  \label{hu111}
\end{equation}
We note that $L_{(\widehat{\mathbf{a}})}$ can also be written as $%
L_{z^{\prime }}$ since as argued in the section on probability amplitudes
and probability densities, it is convenient to think of the vector $\widehat{%
\mathbf{a}}$ as defining a new $z$ direction, denoted by $z^{\prime }$.

It is expected that all generalized spherical harmonics satisfy the
eigenvalue equation, Eq. (\ref{hu110}). This is tedious to prove in
practice, and has not been done for the present case $l=2$. This will be
tackled in the near future, since it is an important part of the proof of
the correctness of the philosophy underlying this work.

\section{References}

1. Mweene H. V., ''Derivation of Spin Vectors and Operators From First
Principles'', quant-ph/9905012

2. Mweene H. V., ''Generalized Spin-1/2 Operators and Their Eigenvectors'',
quant-ph/9906002

3. Mweene H. V., ''Alternative Forms of Generalized Vectors and Operators
for Spin 1/2'', quant-ph/9907031

4. Mweene H. V., ''Spin Description and Calculations in the Land\'e
Interpretation of Quantum Mechanics'', quant-ph/9907033

5. Mweene H. V., ''Vectors and Operators for Spin 1 Derived From First
Principles'', quant-ph/9906043

6. Mweene H. V., Unposted results on spin 3/2 systems.

7. Mweene H. V., ''Generalized Probability Amplitudes for Spin Projection
Measurements on Spin 2 Systems'', quant-ph/0502005

8. Mweene H. V., Unposted results on spin 5/2 systems.

9. Mweene H. V., ''New Treatment of Systems of Compounded Angular
Momentum'', quant-ph/9907082.

10. Mweene H. V., ''Derivation of Standard Treatment of Spin Addition From
Probability Amplitudes'', quant-ph/0003056

11. Mweene H. V., ''Generalized Spherical Harmonics'', quant-ph/0211135

12. Land\'e A., ''From Dualism To Unity in Quantum Physics'', Cambridge
University Press, 1960.

13. Land\'e A., ''New Foundations of Quantum Mechanics'', Cambridge
University Press, 1965.

14. Land\'e A., ''Foundations of Quantum Theory,'' Yale University Press,
1955.

15. Land\'e A., ''Quantum Mechanics in a New Key,'' Exposition Press, 1973.

16. Mweene H. V., ''Proposed Differential Equation for Spin 1/2'', \textit{%
Proc. Third Int. Workshop on Contemporary Problems in Mathematical Physics,
Cotonou 2003}, ed. J. Govaerts, M. N. Hounkounnou and A. Z. Msezane (World
Scientific, 2004), quant-ph/0411060.

\end{document}